\documentclass[prl,amsmath,aps,twocolumn,superscriptaddress]{revtex4}
\usepackage{graphicx}
\usepackage{color}
\begin{document}

\newcommand{\me}{\mathrm{e}}
\newcommand{\mi}{\mathrm{i}}
\newcommand{\dif}{\mathrm{d}}
\newcommand{\nn}{\nonumber}
\newcommand{\be}{\begin{equation}}
\newcommand{\ee}{\end{equation}}
\renewcommand{\epsilon}{\varepsilon}

\newcommand{\fig}[2]{\includegraphics*[width=#1\columnwidth]{./#2}}
\newcommand{\Fig}[1]{\includegraphics*[width=\columnwidth]{./#1}}
\newcommand{\FFig}[1]{\includegraphics*[width=0.85\columnwidth]{./#1}}
\newcommand{\FFFig}[1]{\includegraphics*[width=0.35\columnwidth,angle=270]{./#1}}

\def \bp {\mbox{\boldmath $\partial$}}

\title{The semiflexible fully-packed loop model and interacting rhombus tilings}
\author{Jesper Lykke Jacobsen}
  \affiliation{Laboratoire de Physique Th\'eorique de l'\'Ecole Normale
  Sup\'erieure, 24 rue Lhomond, 75231 Paris, France.}
\author{Fabien Alet}
  \affiliation{Laboratoire de Physique Th\'eorique, Universit\'e de Toulouse,
    UPS, (IRSAMC), F-31062 Toulouse, France}
  \affiliation{CNRS, LPT (IRSAMC), F-31062 Toulouse, France}

\begin{abstract}
  Motivated by a recent adsorption experiment~[M.O. Blunt {\em et al.},
 Science {\bf 322}, 1077 (2008)], we study
  tilings of the plane with three different types of rhombi. An interaction
  disfavors pairs of adjacent rhombi of the same type.
  This is shown to be a special case of a model of fully-packed loops
  with interactions between monomers at distance two along
  a loop. We solve the latter model using Coulomb gas techniques
  and show that its critical exponents vary continuously with the
  interaction strenght. At low temperature it undergoes a
  Kosterlitz-Thouless transition to an ordered phase, which is predicted
  from numerics to occur at a temperature $T\sim110K$ in the experiments.
\end{abstract}

\maketitle

Byzantine artists were probably the first to appreciate the beauty of
tiling the plane with various types of polygons. Nowadays, random
tilings are a major subject in mathematics and theoretical physics,
with many applications to real condensed matter systems.

\begin{figure}
\Fig{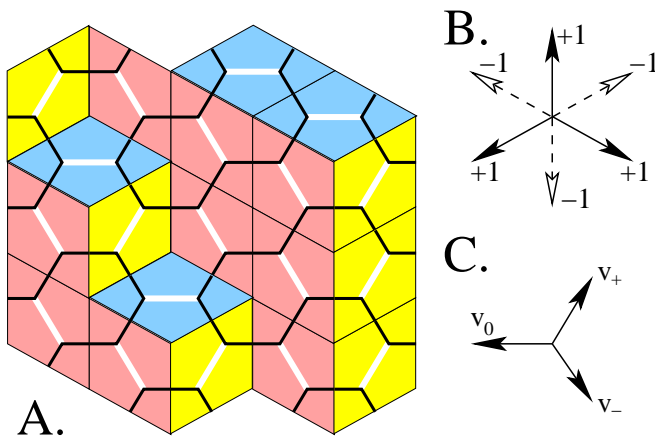}
\caption{(color online) Dimer covering of the hexagonal lattice (thick white lines),
the complementary fully-packed loops (thick black lines), and corresponding
rhombus tiling of the plane (A). Height mapping for the rhombus tilings (B)
and for the fully-packed loop model (C).}
\label{fig1}
\end{figure}

In a very recent experiment, Blunt {\em et al.} \cite{Blunt} study the
adsorption of certain rod-like organic molecules
(para-terphenyl-3,5,3',5'-tetracarboxylic acid) on a graphite substrate
by scanning tunneling microscopy (STM).  An idealization of the
resulting pattern is shown in Fig.~\ref{fig1}A.  The rod-like
molecules (shown as thick white lines) arrange as a dimer covering of
the hexagonal lattice, and interact with their neighbors via hydrogen
bonding of carboxylic acid groups (thick black lines). The bisectors
of the latter define a rhombus tiling with three different types of
rhombi. By image processing the STM pictures, large
tiling configurations could be produced~\cite{Blunt}, with only $\sim 10^{-3}$
defects per molecule at room temperature.

Dimer coverings of various planar lattices have been well studied
theoretically and the corresponding partition functions and various
correlation functions have been obtained exactly
\cite{Dimerold,Blote}. Long-distance behavior is
conveniently analysed in terms of an equivalent height model. For the
hexagonal-lattice case, this is obtained \cite{Blote} by defining a height difference $\pm 1$ to each
displacement along the junction between two rhombi, using the rule in
Fig.~\ref{fig1}B. 
In the continuum limit, one expects
\cite{Blote,Nienhuis1984} the height to behave as a Gaussian free
field with some coupling $g$ which can be computed exactly \cite{Dimerold,Blote}.

The experimentally measured value of $g$~\cite{Blunt} is
however $1.66(8)$ times larger than the theoretical
prediction.  This discrepancy is attributed to interactions between
pairs of neighboring dimers, with an energy penalty $\Delta E > 0$ for a parallel arrangement. This
motivates the study of interacting rhombus tilings (IRT), where each tile
junction carries a weight $w=\me^{-\Delta E/T}$ (respectively $1$) if
it separates rhombi of identical (resp.\ different) types.

The purpose of this Letter is to study a more general model which
contains the IRT as a special case. We define a {\em link} to be any
edge of the hexagonal lattice {\em not} covered by a dimer. In
Fig.~\ref{fig1}A links are shown as thick black lines.  With
appropriate boundary conditions, the links form fully-packed loops,
{\it i.e.}~loops which jointly visit each of the lattice vertices once.
Assign a weight $n$ to each loop. Since each link ${\cal L}_0$ is a
bisector of a tile junction, the weight $w$ must
be attributed to ${\cal L}_0$ if and only if the two links touching
either end of ${\cal L}_0$ are parallel. This defines the partition
function of the semiflexible fully-packed loop (SFPL) model:
\be
 Z_{\rm SFLM} = \sum_{{\rm loops}} n^{N} w^{N_\parallel} \,.
\ee
When $n=1$ we recover the IRT.

Below we solve the SFPL model in the critical regime $-2 < n \le 2$
using Coulomb gas methods \cite{Nienhuis1984,Kondev}. The case $w=1$
was previously shown \cite{Kondev} to be described by two free fields.
We shall see that the interaction $w$ changes the coupling $g_1$ of
one of these fields, leaving the other $g_2$ unchanged. The ratio
$\gamma = g_1/g_2$ is a non-universal decreasing function of $w$.
We compute all critical exponents as functions of $n$ and $\gamma$.
When $w$ becomes smaller than some critical $w_{\rm c}$, the SFPL
model undergoes a Kosterlitz-Thouless (KT) transition to an ordered,
non-critical state. Below we compute analytically the corresponding value of
$\gamma_{\rm c}$, and perform numerical simulations to compute the curve
$\gamma(w)$. This allows to predict the value of the temperature at which
the KT transition should occur in the experiments. We conclude the Letter by a critical discussion
of the experimental realization \cite{Blunt}, and a comparison with
other related loop models.

\smallskip

{\em 2D height mapping.} We briefly review the construction of
Refs.~\cite{Kondev,Kondev1} and show how elements must be modified
to account for the interaction $w$ in the SFPL model.

Orient each loop independently and assign a weight $\me^{-\mi \pi
  e_0}$ to clockwise and $\me^{\mi \pi e_0}$ to counterclockwise
loops. Parametrizing $n = 2 \cos(\pi e_0)$ we recover the correct loop
weight after summing over orientations. Note that this is tantamount
to a weight $\me^{\pm \mi \pi e_0/6}$ to each left (resp.\ right) turn.
Assign a label $v_0$ to edges covered by a dimer, and $v_\pm$ to edges
covered by a link going out of (resp.\ into) a vertex in the even
sublattice. 
Each vertex is then adjacent to three edges, all
carrying different labels $(v_0,v_+,v_-)$. Define the corresponding
two-dimensional vectors ${\bf v}_0 = (-2,0)$ and ${\bf v}_\pm = (1,\pm
\sqrt{3})$ (see Fig.~\ref{fig1}C). Attribute 2D heights ${\bf h}=(h^1,h^2)$ to the dual triangular
lattice, by increasing ${\bf h}$ by ${\bf v}_i$ upon traversing an
edge with label $i$. The traversal must be such that an even (resp.\
odd) vertex is seen on one's left (resp.\ right). The first
component $h^1$ is precisely the 1D height defined by Fig.~\ref{fig1}B
for the rhombus tiling. Being complementary to loops with no orientation,
the rhombi cannot ``see'' $h^2$.

\smallskip

{\em Coulomb gas approach.} The partition function is written as
a functional integral
\be
 Z = \int {\cal D}{\bf h}({\bf x}) \, \exp
 \left( -S[{\bf h}({\bf x})] \right) \,,
\ee
where, by an abuse of notation, ${\bf h}({\bf x})$ denotes the
continuum limit of the height defined above, and the
Euclidian action consists of three terms, $S = S_{\rm E} + S_{\rm B} +
S_{\rm L}$.  The elastic term $S_{\rm E}$ is constrained by rotational
invariance to take the form [with summation
over repeated indices] $ S_{\rm E} = 1/2 \int \! {\rm d}^2{\bf x}  \, g_{{\alpha}{\beta}} \,
 \bp h^{\alpha}\cdot \bp h^{\beta}$,
where $\bp=(\partial_1, \partial_2)$ is the usual gradient. The
$D=2$-dimensional symmetric tensor $g_{\alpha \beta}$ is further
constrained by symmetries. First, since loop orientations are
eventually summed over, the action must be invariant under $v_+ \to
v_-$, viz., $(h^1,h^2) \to (h^1,-h^2)$, implying $g_{12}=0$. We denote henceforth $g_1 \equiv g_{11}$ and $g_2 \equiv
g_{22}$. Second, a cyclic permutation of $(v_0,v_+,v_-)$ maintains the
chirality of the loop turns at each vertex, and is thus a symmetry for
$w=1$.  This implies $g_1  = g_2 $ in this case \cite{Kondev}.

\begin{figure}
\FFig{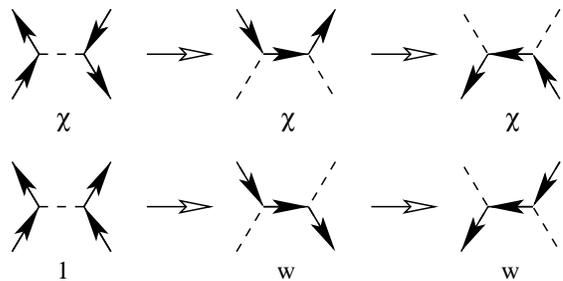}
\caption{The upper group of three configurations for a vertex pair has
  weights $\chi \equiv \me^{i\pi e_0/3}$ which are invariant under the
  cyclic permutation of labels $(v_0,v_+,v_-)$, whereas the weights of
  the lower group differ by factors of $w$.}
\label{fig2}
\end{figure}

\begin{figure}
\Fig{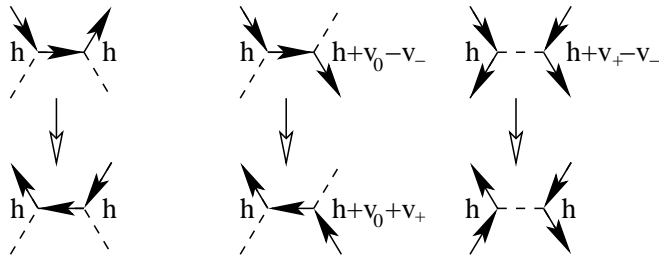}
\caption{Each column shows a pair of configurations with identical link
  positions, but different loop orientations. The middle column comes with
  a weight $w$.}
\label{fig3}
\end{figure}

However, the cyclic permutation is {\em not} a symmetry of the SFPL
model for $w \neq 1$. To see this, we inspect all possible
configurations of the pair of vertices surrounding a fixed edge ${\cal
  E}_0$. By rotation symmetry we can take ${\cal E}_0$ horizontal with
an even vertex on its left end. By reflection symmetry it suffices to inspect six out of twelve configurations. This
gives two groups of three configurations related by the cyclic
permutation. As seen from Fig.~\ref{fig2} the members of the second
group differ by factors of $w$, proving the statement.
We now argue that changing the weight $w$ will modify $g_1$ but leave
$g_2$ unchanged. To this end, we consider pairs of local
configurations having identical link positions, but which differ by
the loop orientation.  In Fig.~\ref{fig3} we show three such pairs
(all others can be found by reflection and rotation) and evaluate the height gradient along the
middle edge. In all cases, the $w$ interaction
must not distinguish between two members of a pair.  We conclude that
$w$ must not couple to ${\bf v}_+ - {\bf v}_- = (0,2\sqrt{3})$, which
is proportional to $h^2$. Conversely, configurations in the middle
column of Fig.~\ref{fig3} have weight $w$ and are the only ones to
have a height gradient along ${\bf v}_0 \propto h^1$. This proves the
claim.

The action also contains a boundary term 
$S_{\rm B} = {\rm i}/(4 \pi) \int \! {\rm d}^2{\bf x} \, ({\bf e}_0 \cdot {\bf h}) {\cal R}({\bf x})$ 
where ${\cal R}$ is the scalar curvature. The background electric
charge ${\bf e}_0$ is easily computed on the cylinder, where it
ensures the correct weighting of non-contractible loops. We have ${\bf
  e}_0 \cdot {\bf v}_0 = 0$ and ${\bf e}_0 \cdot {\bf v}_\pm = \pm \pi
e_0$, implying ${\bf e}_0 = (0,\pi e_0/\sqrt{3})$.

Finally, the Liouville term $S_{\rm L}$ is the continuum limit of the
local vertex weights. The height is compactified \cite{Kondev}
with respect to a triangular lattice ${\cal M}$ of side $2\sqrt{3}$
spanned by ${\bf v}_\pm - {\bf v}_0$. We can therefore expand $S_{\rm
  L}$ as a Fourier series over the vertex operators $\me^{\mi {\bf
    e}\cdot{\bf h}}$, where the electric charges ${\bf e}$ belong to the
lattice ${\cal E}$ reciprocal to ${\cal M}$. It suffices to keep the
most relevant term which has the same periodicity as the vertex
weights.  The inclusion of $w$ does {\em not} change this periodicity,
so we have $S_{\rm L} \sim \me^{\mi {\bf  e}_{\rm s} \cdot{\bf h}}$ where the screening charge reads
${\bf e}_{\rm s} = (0,2\pi/\sqrt{3})$.

\smallskip

{\em Critical exponents.} The critical exponent of an operator ${\cal
  O}_{{\bf e},{\bf m}}$ with electric charge ${\bf e} \in {\cal E}$
and magnetic charge ${\bf m} \in {\cal M}$ reads
\cite{Dotsenko,Kondev1}:
\begin{equation}
 \label{em_dim}
 x_{{\bf e}, {\bf m}} = \left[ e_\alpha(e_\alpha - 2 e_{0\alpha})/g_\alpha + g_\alpha m_\alpha^2 \right]/(4 \pi) \,.
\end{equation}
The corresponding two-point functions decay with distance
$r$ as $r^{-2 x_{{\bf e},{\bf m}}}$. To keep the model critical
we impose \cite{Kondev1} the exact marginality of $S_{\rm L}$,
whence $x_{{\bf e}_{\rm s},{\bf 0}} = 2$, or
\be
 \label{g2}
 g_2 = (1-e_0)\pi/6 \,.
\ee
The other coupling $g_1$ depends non-universally on the microscopic
weight $w$. Henceforth, we express everything in terms of $e_0$ and
the ratio $\gamma \equiv g_1/g_2$. Note that $\gamma(w)$ is 
monotonically decreasing, and $\gamma(1)=1$. Computing the
actual function $\gamma(w)$ would require an exact solution, but it
is doubtful that the SFPL model is integrable. Below, we perform
numerical simulations to obtain this curve in the $n=1$ case relevant to experiments.

The central charge is determined from the background electric charge
as $c=2+12 x_{{\bf e}_0,{\bf 0}} = 2-6 e_0^2/(1-e_0)$, independent of $\gamma$.

An important class of critical exponents is the so-called watermelon
exponents $x_k$. They measure the probability of having $k$ oriented
loop strands emanating from some small neighborhood (of size a few
lattice constants) and absorbed by some other neighborhood at distance
$r \gg 1$. The corresponding height defect (vortex) has magnetic
charge ${\bf m}_k$ which is computed by noting that ${\bf v}_+ - {\bf
  v}_-$ generates a pair of strands, and $2{\bf v}_0- {\bf v}_+$
generates a single strand \cite{Kondev}. Setting $\delta_k = k \mbox{
  mod } 2 \in \{0,1\}$ we find explicitly ${\bf m}_k = (-3
\delta_k,\sqrt{3}k)$. This leads to
\be
 x_k = x_{{\bf e}_0,{\bf m}_k} =
       \left( k^2 + 3 \gamma \delta_k \right) (1-e_0)/8
       - e_0^2/(2-2e_0) \,.
 \label{watermelon}
\ee

Another type of vortex corresponds to having a vertex not
visited by any loop. 
The corresponding magnetic charge is ${\bf m}_T = 3 {\bf v}_0 = (-6,0)$, and the exponent
\be
 x_T = x_{{\bf 0},{\bf m}_T} = 3 \gamma (1-e_0)/2 \,.
 \label{thermal}
\ee

In the IRT model, the second height component is ``invisible'' and only
$x_1$ and $x_T$ are meaningful. In the
tiling picture, the corresponding defects are compounds of three
(resp.\ four) elementary triangles forming a trapezoid (resp.\
triangle) of base length two. Since $e_0 = 1/3$, we have $g_2 =
\pi/9$, $c=1$, $x_1=\gamma/4$ and $x_T = \gamma$.

\smallskip

{\em Kosterlitz-Thouless transition.} Due to the compactification, any
functional of the heights can be expanded over vertex operators with
charges in ${\cal E}$, a triangular lattice of side $2\pi/3$.
The crucial step in solving the SFPL model was to fix the coupling
by requiring the exact marginality of $S_{\rm L}$. By examining the
symmetries of local vertex weights, one finds \cite{Kondev} that vertex operators appearing in the expansion
of $S_{\rm L}$ have charges in a sublattice ${\cal E}_{\rm L} \subset {\cal E}$, a
triangular lattice of side $2\pi / \sqrt{3}$ spanned by the
second-shortest vectors in ${\cal E}$.  For $e_0 > 0$, the most
relevant vertex operator has ${\bf e}={\bf e}_{\rm
  s}=(0,2 \pi/\sqrt{3})$, and the solution Eq.~(\ref{g2}) was obtained by
using this as the screening charge, {\it i.e.}~by setting $x_{{\bf e}_{\rm
    s},{\bf 0}} = 2$.

One can add another term $S_{\rm F}$ to the action that favors domains where
the height interface is locally flat. Its vertex operators have
charges in a sublattice ${\cal E}_{\rm F} \subset {\cal E}_{\rm L}$, a triangular
lattice of side $2\pi$ spanned by the second-shortest vectors in
${\cal E}_{\rm L}$. The interface is in a rough, critical (resp.\ smooth,
non-critical) phase when $S_{\rm F}$ is irrelevant (resp.\ relevant).
Increasing $\gamma$ beyond a certain critical value $\gamma_{\rm c} >
1$ induces a KT transition to the smooth phase. The most relevant vertex operator
in $S_F$ has ${\bf e}_{\rm F}=(\pm 2\pi,0)$. We can find $\gamma_{\rm c}$ by
setting $x_{{\bf e}_{\rm F},{\bf 0}} = 2$, yielding
\begin{equation}
 \gamma_{\rm c} = (1-e_0)/3 \,.
 \label{gammac}
\end{equation}

In the non-critical phase, the couplings $g_1$ and $g_2$ will
renormalize to infinity, corresponding to a microscopic parameter $w
\to 0$. The situation $w=0$
corresponds in Fig.~\ref{fig1}A to the links forming small loops of length six around one
of the three sublattices (denoted ${\cal L}_a$ with $a=1,2,3$) of the
triangular lattice. Which sublattice is selected is a matter of
spontaneous symmetry breaking. Equivalently, in the rhombus picture,
the average number of rhombi $N_a$ touching a vertex of ${\cal L}_a$
will saturate to $(N_1,N_2,N_3) = (6,3,3)$ or any permutation thereof.

The exact exponents at the KT transition are found by inserting
Eq.~(\ref{gammac}) into Eqs.~(\ref{watermelon}) and (\ref{thermal}). Note in
particular that $x_T(\gamma_{\rm c}) = 9/2$ is independent
of $e_0$.

\smallskip

{\em Experimental realization.} As mentioned in the introduction, a
handsome experimental realization of the IRT model appeared
recently~\cite{Blunt}. The energy scales of the experiment are such
that, once formed, the tilings are static (but certain defects can
move around dynamically; see below). Statistics on the height
fluctuations can however be obtained by taking STM pictures of various
regions. These fluctuations were reported to be critical \cite{Blunt},
and an experimental value $\gamma_{\rm exp} = 1.66 \pm 0.08$ was
observed. Since $\gamma_{\rm exp} < \gamma_{\rm c}$ with $\gamma_{\rm
  c}=9/2$ in the $e_0=1/3$ IRT case, our analysis confirms that the
system is indeed critical.

\begin{figure}
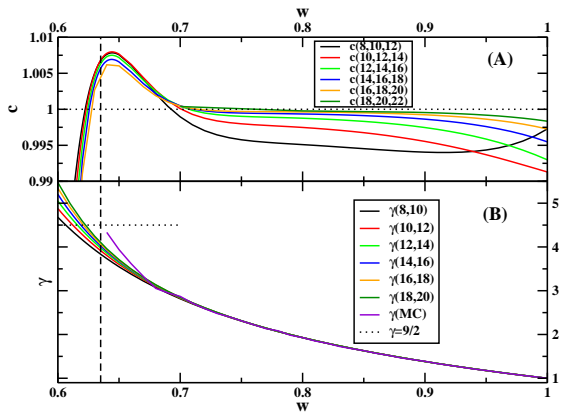

\FFig{fig4.eps}
\caption{(color online) Numerical simulations of the IRT model. As functions of $w$: (A)
  Central charge $c$, obtained from three-point fits of the free energy on
  cylinders of circumference $L$. (B) Coupling constant
  ratio $\gamma$, found from two-point fits for the
  critical exponent $x_1$ (TM data), and from
  winding number fluctuations on a $768 \times 256$ sample (MC data). The
  long-dashed vertical line denotes $w_c=0.635$. 
}
\label{fig4}
\end{figure}

To fix the experimental energy scales within the IRT model, we perform numerical
simulations with Transfer Matrix (TM) and Monte Carlo
(MC) techniques similar to those developped in
Ref.~\cite{Alet}. Fig.~\ref{fig4}A shows $c$ as a function of
$w$, as determined from the TM calculations. A clear $c=1$ plateau appears,
corresponding to the critical phase for $w>w_c$. We
also measured $\gamma$ from the determination of $x_1$ in TM
simulations and winding number fluctuations in MC
simulations~\cite{Alet,Boutillier}. The resulting $\gamma(w)$ curve is
displayed in Fig.~\ref{fig4}B. Extrapolations of $w_{\rm c}$ to the thermodynamic limit
are made: (A) from the TM data by studying the intersections $c=1$ and
$\gamma_{\rm c}=9/2$, giving $w_{\rm c}=0.635(2)$ in both cases,
and (B) from the MC data by styding order parameter fluctuations, giving
$w_{\rm c}=0.640(5)$.
We also find that the experimental value
$\gamma_{\rm exp}=1.66(8)$ corresponds to $w_{\rm exp}=0.845(15)$, allowing to determine the energy scale of
nearest-neighbor interactions as $\Delta E \simeq 4.25$meV. We therefore
predict the KT transition to occur at $T = T_{\rm c} \simeq 110K$ in
the experimental compound. The transition could be observed by monitoring $\gamma_{\rm exp}$ up to the
temperature where it takes the value $\gamma_{c}$, as in
Ref.~\onlinecite{roughening}. For the precise compound of Ref.~\cite{Blunt}, this
will require performing the experiment in vacuum, to avoid that the solvent
freezes \cite{privcomm}.

Among the two possible topological defects ${\bf m}_1$ and ${\bf m}_T$
in the IRT model, the former is by far the most probable, since
$x_T = 4 x_1$ ($= \gamma$). Defects of the ${\bf m}_T$ type, if
observable, would indeed be very closely bound. In the dimer language, the
${\bf m}_1$ defect can correspond either to zero or two dimers incident to the same vertex. Both
possibilities were observed in the STM scans, although the latter
was dismissed as a transient image artifact (see in particular
Fig.~3E of Ref.~\cite{Blunt}). The dynamics of defect pairs should make it experimentally possible to
gather statistics on their relative separation $r$. Given $\gamma_{\rm
  exp}$, the above theory predicts the corresponding power law.

\smallskip

{\em Discussion.} We have solved a model of semiflexible fully-packed
loops, and shown that the bending rigidity couples to just one
of the two coupling constants in the equivalent 2D height model.
Although we have here given a microscopic argument that only $g_1$ was
affected, it should be noted that this is also a consequence of the
field theory.  Indeed, since the screening charge ${\bf e}_{\rm s}$ is
in the $h^2$ direction, $g_2$ is in fact bound to renormalize to the
universal value Eq.~(\ref{g2}). The field theory should remain valid for
other microscopic interactions that have the effect of rendering the height interface stiffer.

The particular case of interacting random tilings obtained when $n=1$
has a physics similar to that of dimer coverings of the {\em square} lattice with local
aligning interactions \cite{Alet}. The SFPL model can also be compared
to the 3D height construction used in Ref.~\cite{Kondev2} to solve the
Flory model of protein melting.

Adding a {\em finite} density of ${\bf m}_T$ type defects to the
SFPL model induces a flow towards the well-known dense phase of the
O($n$) model \cite{Nienhuis82}. Starting from the 2D height mapping,
the $h^1$ component now becomes massive, and the dense phase is
described by $h^2$. Since the corresponding coupling $g_2$ is
insensitive to $w$, we deduce that bending rigidity
is irrelevant in the dense O($n$) model and merely renormalizes
the effective monomer length.

{\em Note added.} Another experimental realization of the IRT model has
appeared very recently~\cite{Han}.

\acknowledgments

We thank the European
Community Network ENRAGE (grant MRTN-CT-2004-005616) and the Agence
Nationale de la Recherche (grant ANR-06-BLAN-0124-03) for support.

\end{document}